\documentclass[preprint,showpacs,preprintnumbers,amsmath,amssymb]{revtex4}
\usepackage{graphicx}
\usepackage{dcolumn}
\usepackage{bm}
\begin{document}
\title{Effect of pinning and driving force on the metastability
effects in weakly pinned superconductors and the determination of
spinodal line pertaining to order-disorder transition}

\author{A. D. Thakur$^{1,~\star}$, S. S. Banerjee$^{2,~\dagger}$, M. J. Higgins$^3$, S. Ramakrishnan$^1$, A. K. Grover$^1$ }
\affiliation{$^1$ Department of Condensed Matter Physics and Materials Science, Tata Institute of Fundamental Research, Mumbai 400005, India \\
$^2$ Department of Physics, India Institute of Technology-Kanpur, Kanpur 208076, India \\
$^3$ NEC Research Institute, Princeton, New Jersey 08540, USA}
\date{\today}

\begin{abstract}
We explore the effect of varying drive on metastability features
exhibited by the vortex matter in single crystals of 2H-NbSe$_2$ and
CeRu$_2$ with varying degree of random pinning. The metastable nature
of vortex matter is reflected in the path dependence of the critical
current density, which in turn is probed in a contact-less way via
ac-susceptibility measurements. The sinusoidal ac magnetic field
applied during ac susceptibility measurements appears to generate a
driving force on the vortex matter. In a nascent pinned single crystal
of 2H-NbSe$_2$, where the peak effect (PE) pertaining to the
order-disorder phenomenon is a sharp first order like transition, the
supercooling feature below the peak temperature is easily wiped out by
the reorganization caused by the ac driving force. In this paper, we
elucidate the interplay between the drive and the pinning which can
conspire to make the path dependent ac-susceptibility response of
different metastable vortex states appear identical. An optimal
balance between the pinning and driving force is needed to view the
metastability effects in typically weakly pinned specimen of low
temperature superconductors. As one uses samples with larger pinning
in order to differentiate the response of different metastable vortex
states, one encounters a new phenomena, viz., the second magnetization
peak (SMP) anomaly prior to the PE. Supercooling/superheating can
occur across both the PE and the SMP anomaly and both of these are
known to display non-linear characteristic as well. Interplay between
the path dependence in the critical current density and the
non-linearity in the electromagnetic response determine the
metastability effects seen in first and the third harmonic response of
the ac susceptibility across the temperature regions of the SMP and
the PE. The limiting temperature above which metastability effects
cease can be conveniently located in the third harmonic data, and the
observed behavior can be rationalized within the Beans Critical State
model. A vortex phase diagram showing the different vortex phases for
a typically weakly pinned specimen has been constructed via the ac
susceptibility data in a crystal of 2H-NbSe$_2$ which shows the SMP
and the PE anomalies. The phase space of coexisting weaker and
stronger pinned regions has been identified. It can be bifurcated into
two parts, where the order and disorder dominate, respectively. The
former part continuously connects to the reentrant disordered vortex
phase pertaining to the small bundle pinning regime, where the
vortices are far apart, interaction effects are weak and the
polycrystalline form of flux line lattice prevails.

\end{abstract}


\maketitle

\section{Introduction}

The richness of the phenomenon of the Peak Effect (PE) in the mixed
state of type-II superconductors has engrossed the vortex physics
community for over forty years \cite{1,2,3,4,5}. The ubiquitous PE
phenomenon is an anomalous enhancement in the critical current density
(J$_c$) of a superconductor as a function of applied field or
temperature in the vicinity of the superconducting to the normal state
transition \cite{1,2}. The vortex matter can be viewed as an elastic
medium in a random pinning environment along with the influence of
thermal fluctuations acting on the pinned vortices. According to a
heuristic argument due to Pippard \cite{6} and subsequent theoretical
work by Larkin and Ovchinnikov \cite{7}, the PE is considered to be
triggered by an incipient softening of the elastic vortex lattice as
the normal state is approached. The softer lattice presumably gets
conformed to its pinning environment more snuggly, thereby producing
an enhancement in the pinning (or J$_c$). The notion that the PE can
be associated with an inevitable and eventual softening of the vortex
lattice encouraged the widespread belief that the PE is a (precursor)
signature of the phenomenon of flux line lattice (FLL) melting
\cite{8,9}. It is useful to recall at this juncture that FLL melting
phenomenon has been unambiguously verified only in the anisotropic
high temperature cuprate superconductors (HTSC) \cite{10}. The PE
phenomenon has, however, been widely studied in low T$_c$
superconductors (LTSC) as well as in the HTSC. In the LTSC, the
smallness of the Ginzburg number, which measures the relative
importance of thermal fluctuations vis-\'a-vis superconducting
condensation energy, makes the thermally driven FLL melting line lie
very close to the H$_{c2}$(T) line \cite{3}. The investigations in a
variety of samples of LTSC reveal that the anomalous behavior
pertaining to the PE could surface up sufficiently below the
H$_{c2}$(T)/T$_c$(H) values and the separation between the onset
position of the PE anomaly and T$_c$(H) line correlates well with the
level of effective disorder in the sample \cite{11}. Such observations
seem to suggest that, generically, the PE as a phenomenon needs a
clarification and distinction from the pristine thermally driven FLL
melting transition.

In any realistic sample of a type II superconductor, the inevitable
presence of residual quenched random disorder is anticipated to
compromise the perfect translational symmetry of the Abrikosov flux
line lattice state. However, it was argued by Larkin and Ovchinnikov
(LO) \cite{12} that in the presence of pinning, the spatial order in
the vortex matter could survive within a domain having dimensions much
larger than the intervortex spacing, but smaller than the typical
sample size. In recent times, Giamarchi and Le Doussal \cite{13} have
shown that the spatial correlations in the weakly pinned vortex matter
could decay much more slowly, in an algebraic manner, such that a good
positional and high orientational order, equivalent to a quasi-long
range order (notionally a `Bragg glass' state), can be observed over
length scales comparable to the sample dimensions. A contemporary view
that has gained acceptance about the PE anomaly is that it represents
a transition from a weakly pinned Bragg glass state to a stronger
pinned multi (or micro) domain vortex glass state. Bulk ac
susceptibility studies in a very weakly (nascent) pinned
superconducting specimen of 2H-NbSe$_2$ had pointed towards the
association of the PE with the first order nature of the underlying
transition in the pinned vortex matter \cite{14}. Eventually, the
local ac susceptibility measurements by scanning micro-Hall bar
microscopy \cite{4} in the typically weakly pinned single crystals of
2H-NbSe$_2$ directly elucidated the presence of an interface
separating the weaker and stronger pinned regions across the peak
effect region, thereby endorsing its first order nature and the
associated metastability effects.

The presence of a first order transition line in the magnetic phase
(H, T) diagram of a weakly pinned superconductor imbibes in it the
notion of supercooling and superheating effects across it
\cite{15,16,17}. An STM imaging study \cite{18} alongwith an analysis of
the instantaneous positions of the individual vortices in a very
weakly pinned crystal of 2H-NbSe$_2$ led to the surmise that
neighboring vortices execute collective motion below the onset
temperature of the PE, which transforms to positional fluctuations
(random excursions) of individual vortices above it. A small angle
neutron scattering (SANS) study \cite{19} in a typically weakly pinned
crystal of Nb could reveal the superheating of the ordered Bragg glass
phase above the onset temperature of the PE alongwith the usual
supercooling of the disordered vortex glass phase below it. An
elucidation of the superheating effects across the PE in a variety of
single crystals of 2H-NbSe$_2$ by bulk transport measurements using a
fast current ramp procedure \cite{20} has resulted in the demarcation
of a generic spinodal line corresponding to the limit of the
superheating of the ordered state. Above the spinodal line, the
threshold force needed to depin and drive the lattice (i.e., critical
current density J$_c$) is independent of the thermomagnetic history of
the underlying pinned vortex matter.

The work of Xiao {\it et al.}  \cite{20} on the spinodal line in
conjunction with the correlation between the effective disorder, the
structure across the PE and the metastability effects \cite{14,17,21}
raises an issue relating to the possible connection between the
quenched random disorder and the details characterizing the spinodal
nature. In the present report, we shall focus on the results
pertaining to this issue. It had been shown earlier \cite{14,17,21}
that with the gradual enhancement in the quenched random pinning, the
first order nature of the order-disorder transition associated with
the PE is not severely compromised; only the process of disordering
starts to comprise several distinct steps. In a nascent pinned sample
\cite{14}, the interface separating the weaker and stronger pinned
regions is perhaps very fragile, and the metastability effects in such
a sample get wiped out easily by an infinitesimally small driving
force. In typically weakly pinned specimen of LTSC systems (like, Nb
crystal used in SANS experiment \cite{19}, or the 2H-NbSe$_2$ crystals
used in micro-Hall bar microscopy \cite{4}), the metastable states are
more robust and the thermomagnetic history effects can be conveniently
probed via contactless ac susceptibility measurements, with a minimal
level of an ac driving force. By performing broad band noise
measurements in the in-phase ac susceptibility ($\chi^\prime_\omega$)
and the third harmonic ($\chi^\prime_{3\omega}$) measurements, one can
demarcate the limits of metastability and determine the spinodal line
in the vortex phase diagram. We elucidate the above stated behavior in
the samples of two very different superconducting compounds \cite{21},
viz., 2H-NbSe$_2$ (T$_c$ $\sim$ 6 K) and CeRu$_2$ (T$_c$ $\sim$ 6.3 K)
to establish the generic nature of the spinodal line, so determined.

\section{Experimental}

The ac susceptibility studies have been carried out using a
conventional double coil arrangement affixed coaxially inside a
superconducting magnet coil and employing a mutual inductance bridge
\cite{22}. Most of the data have been recorded in the frequency interval of
21 Hz to 211 Hz and with an ac amplitude (h$_{ac}$) lying in the range
of 0.5 Oe to 3.5 Oe (r.m.s.). The samples studied include the crystals
X, Y$^\prime$, Y, Z and Z$^\prime$ of 2H-NbSe$_2$ \cite{14,17,23} and
a single crystal of CeRu$_2$ \cite{21}. The crystal X of 2H-NbSe$_2$
is the most weakly pinned sample, with a T$_c$(0) = 7.225 K, the
crystals Y$^{\prime}$ and Y are somewhat stronger pinned than the
crystal X with T$_c$(0) = 7.25 K and 7.17 K, respectively. The
crystals Z and Z$^\prime$ have T$_c$(0) values in the neighborhood of
$\sim$ 6 K and they probably contain few hundred ppm of Fe impurity in
them \cite{21}. From the sample X to Z$^\prime$, the J$_c$ values at
4.2 K in the low field region ($\sim$ 1kOe) vary from about 10
A/cm$^2$ to 10$^3$ A/cm$^2$ \cite{23}. We reckon that the FLL
correlations in the vortex matter in the crystals Z and Z$^\prime$ of
2H-NbSe$_2$ are like those in the typically weakly pinned crystals of
other low T$_c$ superconducting compounds, e.g., CeRu$_2$,
Ca$_3$Rh$_4$Sn$_{13}$, etc. \cite{21,24}.

\section{Results and Discussions}

\subsection{AC susceptibility measurements in the crystals of
2H-NbSe$_2$ with varying pinning}

Figure 1 shows the in-phase ac susceptibility response,
$\chi^\prime_\omega$, in the different crystals X, Y$^{\prime}$, Y, Z and
Z$^\prime$ of 2H-NbSe$_2$ in a field of 15 kOe. The ac susceptibility
response varies as \cite{25}, $\chi^\prime_\omega \sim - 1 + \alpha
h_{ac}/J_c$, where $\alpha$ is a geometry and size dependent
factor. This implies that the $\chi^\prime_\omega$ response closely
mimics the behavior of the J$_c$ in the sample. In different samples,
we compare the J$_c$ values of the vortex states corresponding to
different thermomagnetic histories, viz., the zero field cooled (ZFC)
state and the two field cooled (FC) modes. The response of a FC state
is first measured while cooling down (FCC), and later measured while
warming up (FCW).

Figure 1(a) shows the $\chi^\prime_\omega$ response in the cleanest
crystal X for the vortex states in the ZFC, FCC and FCW modes. All the
three responses are observed to overlap, such that no distinction can
be made amongst them.  In this crystal, we find the PE to be a very
sharp feature (which occurs over a width of 20-40 mK or
so). Considering that the crystal X has the least amount of disorder,
one may argue that the properties of the first order nature of the PE
transition should get best exemplified in this crystal. It was,
therefore, expected that a substantial hysteresis in the
$\chi^\prime_\omega$ responses should have been evident between the
ZFC and the FCC states due to the possible supercooling effect below
the onset temperature of the first order transition. While the data
shown in figure 1(a) for the vortex states in a field of 15 kOe is a
representative one for the crystal X, we found the absence of any
thermomagnetic history dependence in the $\chi^\prime_\omega$ response
(i.e., differences between the ZFC and the FCC or the FCW states)
across the PE transition at all magnetic fields in this sample (all
data not being shown here). A possible reason for the absence of the
hysteresis between the ZFC and the FC states in the sample X is a
subtle effect of the superimposed ac magnetic field h$_{ac}$. We
elucidate next an additional facet of h$_{ac}$ in the ac
susceptibility measurements on the pinned vortex matter, which is
different from its obvious role in measuring the shielding response
from the macroscopic screening currents induced by its imposition on
the sample.

Figure 1(b) displays the $\chi^\prime_\omega$ response of the vortex
state with two different thermomagnetic histories, viz., the ZFC and
the FCW, in the crystal Y$^{\prime}$ at 15 kOe. We observe that the
sharpness of the PE in this crystal is comparable to that in the
crystal X, but the response of the ZFC and FC vortex states can be
made to differ, unlike the situation observed in the crystal X. We
find that for the ZFC and FCW modes, the $\chi^\prime_\omega$
responses in the crystal Y$^{\prime}$ coincide if the h$_{ac}$ is kept
`on' while the sample is initially cooled down in a dc magnetic field
to the lowest temperature (which was 4.3 K in the present case). Such
a situation is identical to that in the sample X. However, if the
sample Y$^{\prime}$ is field cooled with the h$_{ac}$ switched off
during the cooling, and it is switched back on only when one begins
the FCW measurement, then one observes a measurable difference in the
ZFC and the FCW responses. This behavior is indeed different from that
observed in the crystal X. These imply that depending on the pinning
in the sample, h$_{ac}$ has the potential to modify the path
dependence in J$_c$. Such an attribute becomes further exemplified as
we examine the results on the samples with higher pinning, viz., the
crystals Y and Z$^\prime$. In these samples, it is significant to note
that the differences in the responses of the ZFC and the FCW states
survive, whether the h$_{ac}$ is kept {\it on} or {\it off} during the
initial field cooling procedure (see figures 1(c) and 1(d)). {\it Note
that the limiting temperature below which the $\chi^\prime_\omega$
response exhibits thermomagnetic history dependence (cf. the positions
of arrows marked in the different panels of figure 1) could be a
function of h$_{ac}$ and the extent of effective pinning in the
sample}. To emphasize another observation, the notional peak position
of the PE in the ZFC (or the FCW) mode does not necessarily mark the
limiting temperature above which the thermomagnetic history effects in
the $\chi^\prime_\omega$ response cease.

>From the $\chi^\prime_\omega$ responses in figures 1(c) and 1(d), it
is evident that J$_c^{zfc}$ $<$ J$_c^{fcw}$. To understand the
behavior of J$_c$ corresponding to the vortex states with different
thermomagnetic histories, we take recourse to a contempory view
\cite{4} that in the PE region the vortex state comprises {\it an
admixed inhomogeneous phase}, with the coexistence of ordered and
disordered regions. The ZFC state prior to the PE represents an
ordered weakly pinned state, which is characterized by a low J$_c$
value. The FCW state on the other hand could correspond to the phase
above the peak temperature, viz., a predominantly disordered, strongly
pinned vortex state that can get supercooled down to the low
temperatures during field cooling. The FCW state, therefore, carries a
higher J$_c$ value. We have elucidated through the results in figure
1(b) that the non-observation of metastable phases in the
$\chi^\prime_\omega$ response need not imply the absence of path
dependence in J$_c$.

In the presence of very weak pinning, the metastable states of the
vortex matter could be very fragile. Any attempt to couple the system
to the external environment perturbs the underlying vortex matter such
that it is driven into a different state, thereby masking (i.e.,
altering) the pristine information pertaining to the
metastability. The process of keeping h$_{ac}$ switched `{\it on}'
during field cooling the vortex matter can also help the disordered
regions in exploring the possibilities of transformation into the
ordered regions of the ZFC mode. This is best exemplified by the data
in figure 1(a), wherein in the crystal X, the $\chi^\prime_\omega$
response is identical, irrespective of whether the h$_{ac}$ was kept
`{\it on}' or `{\it off}' during initial field cooling. In the sample
X, even while h$_{ac}$ was kept {\it off} during FC, the application
of a small h$_{ac}$ to measure the $\chi^\prime_\omega$ response
during the FCW mode resulted in driving the field cooled state towards
the ordered state of the ZFC mode.

In the sample Y$^{\prime}$, where the pinning is somewhat stronger than
that in sample X, the option of keeping the h$_{ac}$ switched {\it
on/off} during the initial field cooldown procedure demonstrates the
competition between progressive enhancement in pinning and the
annealing effect of h$_{ac}$. The enhanced pinning in sample
Y$^{\prime}$ arrests the hysteresis across the PE and allows the
difference between the ZFC and the FCW state to be observed clearly
before the onset temperature of the peak effect (T$_p^{on}$). The
inevitable dynamical changes which accompany the PE phenomenon make
the $\chi^\prime_\omega$ responses for the ZFC and the FCW states
nearly identical for T $>$ T$_p^{on}$ in the sample Y$^{\prime}$. The
progressive increase in the pinning can therefore aid the
metastability of the supercooled (FC) phase against the thermal
fluctuations and the possible perturbations from h$_{ac}$. To search
for the true location, where the path dependence in J$_c$ ceases, in
other words to determine the spinodal temperature (T$^*$) which is not
influenced by the driving forces, one needs to resort to a sample with
optimal pinning. {\it Therefore, to generate metastability in the
vortex matter and to sustain its imprint in a measurement, one
requires a crucial balance between the strength of the driving force
(h$_{ac}$/transport current), disorder/pinning and the
temperature}. At this juncture, it may also be pertinent to recall and
compare the results of the electrical transport data of Xiao {\it et
al.} \cite{26} in a crystal of 2H-NbSe$_2$ with $\chi^\prime_\omega$
response as in our samples Y and Y$^{\prime}$. It has been reported that
the transport J$_c$ value for the FC state in a very weakly pinned
specimen differs from that in the ZFC state only during the first
ramp-up of the current while recording the I-V data. The J$_c$ values
determined during the ramp-down cycle or during the subsequent ramp-up
cycles were found to be equal to that for the ZFC state. The passage
of transport current presumably reorganizes the disordered state of
the FC mode towards the ordered ZFC mode, when the pinning effects are
in the nascent stage. Xiao {\it et al.}  \cite{20} had to later adopt
the procedure of fast current ramp to explore the metastable states
above the notional onset temperature of the PE. They found that the
superheating of the ordered state could be observed upto a limiting
temperature T$^*$, which exceeds the notional peak temperature of the
PE for the vortex state prepared in the ZFC mode \cite{20}.

The plots in figure 1(d) appear to imply that the metastability
effects in the $\chi^\prime_\omega$ response would cease above the
peak temperature T$_p$, as was conventionally believed
\cite{9,27,28}. The FC state in this sample can be ascribed to a
freezing in of the disordered vortex state at the peak position of the
PE, where the lattice was believed to be homogenously amorphous
\cite{21,28}. However, the recent work by Xiao
{\it et al.}  \cite{20} desires a serious revision of the above presumption. A
transformation that sets in at T$_p^{on}$ results in an admixed state of
ordered and disordered phases. Both these fractions carry different J$_c$
values and possess different temperature dependences. As the
temperature increases from T$_p^{on}$ towards T$_p$, the fraction of the
ordered vortex phase decreases and that of the disordered phase
increases, correspondingly the observed J$_c$ value also increases. The
relative fractions of these two phases and their temperature
dependences determine the peak position (i.e., T$_p$/H$_p$) of the PE
anomaly. In view of this, the relationship, if any, between T$_p$ and the
limiting temperature at which the metastability effects cease is not
very apparent. We have set out to clarify this issue via the ac
susceptibility measurements.

\subsection{Comparison of AC susceptibility data in the crystals of
2H-NbSe$_2$ and CeRu$_2$}

As stated earlier, in order to study the features associated with a
first order transition one should in principle attempt to select a
sample with the least amount of disorder. However, the data in figure
1 show that to study features related to the spinodal temperature
T$^*$ via the ac susceptibility measurements, we were compelled to
choose the samples of Z and Z$^\prime$ category, which have
significantly higher level of disorder than the sample X. While it is
clear that with an enhanced pinning, the metastability features of the
vortex matter stand preserved, it is not however apparent as to
whether the first order character of the PE transition stands retained
at comparatively higher levels of disorder. A trend which emerges from
the $\chi^\prime_\omega$ response in figures 1(a) to 1(d), is that
between T$_p^{on}$ and T$_p$, the PE phenomenon undergoes a broadening
with progressive increase in the effective pinning. It could be argued
that this indicates that a sharp first order like PE transition is
being transformed into a continuous transition.  However, in this
context, it is useful to recall that a novel notion of (multi-step)
fracturing had been demonstrated earlier \cite{21,24}. The first order
nature of the PE transition is not destroyed in samples with higher
pinning, it only transforms into multiple small first order like
jumps, which comprise the notion of fracturing.

Figure 2 shows a comparison of the ac susceptibility response at
different ac amplitudes in a crystal Z of 2H-NbSe$_2$ with that in a
crystal of CeRu$_2$ (cubic system with T$_c(0) \approx$ 6.3 K), with a
level of pinning comparable to that in the crystal Z \cite{21}. In
both these samples, we observe in figures 2(a) and 2(c) that although
the PE is a broad feature, yet it is characterized by multiple sharp
jumps commencing at T$_p^{on}$ in the $\chi^\prime_\omega$(T)
response, characterizing the fracturing transition. A well ordered
vortex phase is considered to transform into a multi-domain like
state, with pockets of ordered and disordered regions created,
possibly via the progressive permeation of topological defects like
dislocations, into the ordered vortex phase. At each first order like
jump, some particular pockets with a collection of well ordered
vortices (with a pocket characterized by a low J$_c$ value), transform
into a disordered phase with a higher value of J$_c$, locally. More
the number of such domains, with intervening disordered vortex phase,
the average J$_c$ of the sample increases leading to a PE like feature
above T$_p^{on}$. The coexisting ordered and disordered phases are not
nucleated by merely supercooling the vortex matter below a first order
transition, they can also get generated due to the staggered nature of
the first order transition as a consequence of the fracturing
phenomenon. The fraction of domains with disordered vortices is larger
while field cooling below the fracturing transition, and this result
in prominence of the metastability effect.

The above description finds an echo in the plots contained in the
figures 2(b) and 2(d). We observe that in the samples of 2H-NbSe$_2$
and CeRu$_2$, there is a substantial difference between the ZFC and
the FCW states for h$_{ac}$ = 0.5 Oe (cf. figures 2(a) and 2(c)). For
larger h$_{ac}$ values (i.e., between 2 Oe and 3.5 Oe, as shown in the
plots for figures 2(b) and 2(d), the PE becomes a narrower transition
and the fracturing features can get suppressed and, consequently the
metastability response (as reflected by the differences between the
ZFC and FCW responses) also stands reduced. The connection between the
weakening of the fracturing transition and the decrease in
metastability is presumably related to the annealing effect of the
h$_{ac}$ drive in modifying the relative fractions of the ordered and
disorder regions between T$_p^{on}$ and T$_p$. With a larger h$_{ac}$
drive used for $\chi^\prime_\omega$ measurements, the residual
disordered pockets in the ordered vortex matter get transformed into
the ordered ones, prior to the onset of the PE. The vortex matter thus
approaches a single domain like picture of the ordered vortices as in
the samples X and Y$^{\prime}$ of 2H-NbSe$_2$, which then undergoes a
sharp PE transition, rather than a multi-step fracturing
transition. It is also pertinent to note that with the enhancement in
the ordered fraction of vortices in presence of a driving force, the
disordered domains of vortices can permeate and survive in the midst
of the ordered vortex matter only at relatively higher temperature
(cf. the T$_p^{on}$ and T$_p$ values in the figures 2(b) and 2(d) with
those in the figures 2(a) and 2(c), respectively). It is also
instructive to compare FCW responses in the figures 2(a) and 2(c) with
those in the figures 2(b) and 2(d), respectively. The supercooled
state in the FCW mode need not undergo order-disorder transition
across the PE region; such a behavior is evident when h$_{ac}$ is
small (cf. figures 2(a) and 2(c)). However, a larger h$_{ac}$ drive
can reorganize the initial FCC state during warm-up and eventually
depict an order-disorder transition across the temperature region of
PE, as evident in figures 2(b) and 2(d).

Having explained the metastability features associated with the
fracturing transition seen in samples with higher pinning, and
determining the optimum range of parameters for h$_{ac}$, disorder and
temperature, we now dwell on the demarcation of the limit of
metastability in the vortex phase diagram. Figure 3 shows the
$\chi^\prime_\omega$(T) response at different applied fields for ZFC,
FCC and FCW modes in the sample Z$^\prime$ of 2H-NbSe$_2$. Note that
at the lowest temperature, the FCW and FCC responses are identical, at
higher temperatures the FCW state is more ordered in comparison to the
FCC state. Apart from the behavior of metastability and the PE, we
observe an additional anomaly in the behavior of the
$\chi^\prime_\omega$ response on the ZFC branch. In samples having a
level of pinning as in the crystal Z$^\prime$, we can notice the
occurrence of an anomalous enhancement in J$_c$ deep in the mixed
state beginning at T$_{smp}^{on}$, corresponding to the second
magnetization peak (SMP) feature observed in the isothermal
magnetization hysteresis (M-H) loops well before the onset of the
usual PE at T$_p^{on}$ \cite{29}. The SMP feature is different from
the fracturing phenomenon which gets triggered at (or near) T$_p^{on}$
(see figure 3(a)). It suffices to recall and state here that the (two)
anomalies of the SMP and the PE are distinct and different
\cite{29}. It may be worthwhile now to ask as to at what stage does
the phase coexistence of ordered and disordered phase commence, in the
presence of a SMP anomaly. A conventional notion as alluded to in the
description earlier would imply that the phase coexistence region
commenced from the onset position of the PE. We shall now show that
such a notion desires a revision in view of the data presented in the
panels of figure 3. Closely associated with the notion of phase
coexistence and the first order nature of the PE is that of the
metastability.

While examining the metastability response, it is instructive to focus
on the interesting behavior illustrated in the data recorded at 6 kOe
(cf. figure 3(b)). Note that the peak temperature of the PE is
different for different thermomagnetic histories, the highest value
being in the ZFC state, T$_p^{zfc}$. Similar trend can be noted at
other magnetic fields (cf. figures 3(c) and 3(d)). Such a behavior for
the $\chi^\prime_\omega$ response for the vortex states with different
thermomagnetic histories, is analogous to that reported for the
transport critical current I$_c$(T) data in a crystal of 2H-NbSe$_2$
by Xiao {\it et al.}  (cf. figure 1 of Ref.\cite{20}). Vortex states
with different histories are characterized by the different fractions
of the ordered and the disordered phases. The observation that
T$_p^{fcc}$ $<$ T$_p^{fcw}$ $<$ T$_p^{zfc}$ implies that the limiting
temperature at which the sample is homogenously filled with the
disordered phase does not lie in the temperature interval, from
T$_p^{fcc}$ to T$_p^{zfc}$.

The $\chi^\prime_\omega$(T) response for a given H is essentially
dictated by J$_c$(T). The temperature above which
$\chi^\prime_\omega$(T) (or, J$_c$(T)) becomes independent of the
thermomagnetic history of the specimen appears to lie even above the
(highest) peak temperature, i.e., at T $>$ T$_p^{zfc}$. Figure 4(a)
focuses attention onto the plots showing ($\chi^{\prime zfc}_\omega -
\chi^{\prime fcw}_\omega$) and ($\chi^{\prime zfc}_\omega -
\chi^{\prime fcc}_\omega$) at 12 kOe in the sample Z$\prime$. In this
panel one can identify the limiting temperature T$^*$ above which the
$\chi^\prime_\omega$ response becomes path independent, viz., where
the differences ($\chi^{\prime zfc}_\omega - \chi^{\prime
fcw}_\omega$) and ($\chi^{\prime zfc}_\omega - \chi^{\prime
fcc}_\omega$) vanish. In the sample Z$\prime$, due to the relatively
strong pinning and the fracturing phenomenon, the determination of the
limit of metastability effects is insensitive to perturbations from
h$_{ac}$. We believe that the above procedure yields a reasonable
estimate of the spinodal temperature, T$^*$, for 2H-NbSe$_2$ samples
with J$_c$ ($\sim$ 1 kOe, 4.2 K) $\geq$ 1000 Amps/cm$^2$. However, the
above procedure may not suffice for all weakly pinned samples, like,
the crystals X and Y$^\prime$ of 2H-NbSe$_2$. It is therefore
desirable to explore an alternative way to determine the value of
T$^*$.

\subsection{Spinodal temperature T$^*$ and the third harmonic ac
susceptibility measurements}

The critical state model (CSM) relates the macroscopic J$_c$ to the
hysteretic magnetization response of a superconductor \cite{30,31}. It
is well documented \cite{32,33,34,35,36} that new features get added
to the pristine relationship between the J$_c$(H) and the
magnetization hysteresis, when J$_c$(H) does not remain single valued
function of H. For instance, the minor hysteresis loops can display
anomalous characteristics, like, an asymmetric shape \cite{34,35},
excursions beyond the envelope M-H loop \cite{36}, etc., when the
J$_c$(H) turns path dependent. In the back drop of these observations,
it is instructive to examine the response of the third harmonic of the
ac susceptibility across the SMP and PE regions, i.e., from below the
onset temperature of the first of the two anomalous variations in
J$_c$ upto the irreversibility temperature (T$_{irr}$), where the
(bulk) J$_c$ ceases. To be specific, consider the cool down of a
weakly pinned type-II superconducting sample from above a given
T$_c$(H) value to below its T$_{irr}$(H), where the finiteness of
J$_c$ would result in a non-linear magnetization response which could
generate a measurable third harmonic signal in the ac susceptibility
measurements. Such a third harmonic signal would be expected to follow
the increase in J$_c$(T) for a given H as (T$_{irr}$ - T) increases,
as per a prescription of the CSM for path independent J$_c$(H, T)
\cite{30,31}. The onset of the history dependence in J$_c$(H) could
compromise the above stated notion, arising from the applicability of
the CSM. Below T$^*$, due to the onset of the path dependence in
J$_c$, qualitative changes could occur in the behavior of the
nonlinear response at the nucleation of an inhomogeneous phase.

The plots in figures 4(b) to 4(f) explore the limit of the path
dependence in J$_c$(H) via the study of the temperature dependence of
the third harmonic ac susceptibility data, $\chi^\prime_{3\omega}$, in
the sample Z$^\prime$ of 2H-NbSe$_2$. In figure 4(b), we draw
attention specifically to the behavior of $\chi^\prime_{3\omega}$ just
above T$^*$. Warming up from the low temperature side, as the
temperature crosses the limit T$^*$, $\chi^\prime_{3\omega}$(T) is
seen to monotonically decrease and eventually vanish at the
irreversibility temperature, T$_{irr}$ ($<$ T$_c$(H)). There does not
appear any simple correspondence between the $\chi^\prime_\omega$(T)
and $\chi^\prime_{3\omega}$(T) for T$_{smp}^{on}$ $<$ T $<$ T$^*$
(cf. figures 3(d) and 4(b)). Figure 3(d) shows that the
$\chi^\prime_\omega$(T) monotonically decreases above T$_p^{zfc}$,
reflecting the collapse in J$_c$ above the peak position of the
PE. The $\chi^\prime_{3\omega}$(T), on the other hand in figure 4(b),
appears to enhance between T$_p^{zfc}$ and T$^*$. The
$\chi^\prime_{3\omega}$(T) response is seen to turn around only above
T$^*$, and follow the J$_c$(T) thereafter. Thus, the onset of an
inhomogeneous phase at T $<$ T$^*$, produces modulations in the
behavior of nonlinearity or in the $\chi^\prime_{3\omega}$ response.

To establish the assertion on the imprint of the limit of the path
independence in J$_c$ in the $\chi^\prime_{3\omega}$(T) data, we can
examine the above stated behavior at different fields in figures 4(c)
to 4(f), and note one to one correlation between values of T$^*$
determined from $\chi^\prime_\omega$(T) data in different
thermomagnetic histories and the limiting temperatures above which
$\chi^\prime_{3\omega}$(T) monotonically decreases. The T$_p^{on}$,
T$_p^{zfc}$ and T$^*$ values have been marked for each of the curves
in figure 4. At T $<$ T$^*$, the modulations in
$\chi^\prime_{3\omega}$(T) display complex behavior, dictated by the
(intrinsic) changes in the relative fraction of the ordered and
disordered phases (path dependence in J$_c$(T)) and the additional
changes induced by the h$_{ac}$ value in which the
$\chi^\prime_{3\omega}$(T) data are recorded. The T$^*$, however, does
not appear to vary in any noticeable manner with the amplitude of
h$_{ac}$ (all data not shown here). It is also worthwhile to note that
the modulation in the $\chi^\prime_{3\omega}$(T) signal ceases close
to T$_{smp}^{on}$. In fact, if we approach T$_{smp}^{on}$ from the
lower temperature side, the $\chi^\prime_{3\omega}$(T) plots appear to
show an enhancement in the non-linear response at T $>$
T$_{smp}^{on}$. Such an observation would get rationalized by invoking
the notion that the SMP and the PE are non-linear phenomena
\cite{9}. In the absence of complications induced by (possible)
phase-coexistence and history effects, the $\chi^\prime_{3\omega}$(T)
would have been expected to follow the modulation as displayed by
$\chi^\prime_\omega$(T). We believe that the possibility of
phase-coexistence commences at T$_{smp}^{on}$ and lasts upto T$^*$.

\subsection{Correlation between $\chi^\prime_{3\omega}$(T) and
measurement of noise in $\chi^\prime_\omega$(T)} 

To further substantiate that the values of T$^*$ determined from the
$\chi^\prime_{3\omega}$ response indeed physically correspond to the
spinodal temperatures, we explore the correlation between the
$\chi^\prime_{3\omega}$(T) and the noise signal in
$\chi^\prime_\omega$(T) \cite{21}, which can be easily recorded using
a Lock-in amplifier having a wide band filter option. The above stated
noise signal is believed to measure the fluctuations in
$\chi^\prime_\omega$(T) and it is argued \cite{21} to reflect the
possibility of transformations amongst metastable states accessible
from a given mode (ZFC/FC).

Figure 5 depicts the plots of $\chi^{\prime zfc}_{3\omega}$(T) (in
h$_{ac}$ = 3.5 Oe (r.m.s.)) and the noise in $\chi^{\prime
zfc}_\omega$(T) in h$_{ac}$ = 0.5 Oe (r.m.s.) in a field of 5 kOe in
the sample Z of 2H-NbSe$_2$ and in field of 13.5 kOe in the crystal of
CeRu$_2$. In different panels of figure 5, we have marked the
respective values of T$^*$ as determined from the onset of
nonmonotonic modulations in $\chi^\prime_{3\omega}$ response. In view
of the notion of the existence of a homogenously disordered phase
above T$^*$, it is indeed not a coincidence that the noise signals in
the panels (b) and (d) of figure 5 recede to the background value as T
$\to$ T$^*$. Taking cue from earlier studies of noise in transport
experiments \cite{37,38}, Banerjee {\it et al.}  \cite{21} had argued
that the increase in noise at T = T$_p^{on}$ (equivalent to the
temperature of the onset of plastic flow, T$_{pl}$, in Ref. \cite{21})
reflects the possibility of enhancement in transformations amongst the
coexisting \cite{4} metastable states in a fractured (partially
disordered) vortex solid (which exists at T $<$ T$^*$). The setting in
of the decrease in the noise signal as T $\to$ T$_p$ was considered to
imply the effect of phase cancellation of a large number of incoherent
fluctuations as the vortex matter moves towards the fully disordered
state. In such a framework, the possibility of coexistence of ordered
pockets embedded in the disordered medium would cease as T $\to$
T$^*$, and the noise signal would be expected to recede to the
background value. In the context of our present results, the T$^*$(H)
values indeed represent the notion of the spinodal line, T$_s$(H)
\cite{20,39}.

\subsection{Magnetic Phase Diagram in a typically weakly pinned crystal of
2H-NbSe$_2$}  

To comprehend a variety of results presented above, we have
constructed a field-temperature (H, T) diagram for the sample
Z$^\prime$ of 2H-NbSe$_2$ in figure 6. It includes the values of the
onset temperatures of the SMP and the PE anomalies, along with the
values of T$_p^{zfc}$, T$^*$(H), T$_c$(H) and H$_{plateau}$(T). In
sample Z$^\prime$ (or Z), there is adequate perceptible difference in
T$_p^{zfc}$ and T$^*$(H), and this difference enhances as H
increases. However, in the samples Y and Y$^{\prime}$ of 2H-NbSe$_2$,
where the PE manifests as a sharp transition, T$^*$ values lie in very
close proximity of the corresponding peak temperature values. In the
vortex phase diagram of figure 6, the plateau line, H$_{plateau}$(T),
passes through the limiting fields \cite{11,17} below which the
collective pinning regime gives way to the small bundle pinning
regime. It seems appropriate to identify the (H, T) space between the
H$_{plateau}$(T) and the T$_{smp}^{on}$(H) as the Bragg (elastic)
glass region \cite{13}. Above the T$^*$(H) line, where the
metastability effects cease, the vortex matter exists in the pinned
amorphous phase. The phase diagram demarcates the regime of
phase-coexistence \cite{40}. We propose that T$_{smp}^{on}$(H) line
corresponds to the boundary where pockets of disordered phase start
proliferating in the ordered vortex state, however, the fraction of
the ordered phase exceeds that of the disordered phase upto
T$_p^{on}$. Above T$_p^{on}$, the balance starts to rapidly tilt in
favor of the disordered regions and the strongly pinned phase starts
determining the overall electromagnetic response. Above the spinodal
line T$^*$(H), the phase coexistence and metastability features cease
and the sample is homogenously filled with the disordered regions. The
bifurcation of the coexistence phase space into regions I and II,
where the order and the disorder dominate, respectively, finds
additional support from the temperature dependence of
$\chi^\prime_\omega$(T) in the FCW mode in figures 3(c) and (d). Both
sets of data show that J$_c$(T) values for the stronger pinned vortex
solid continue to decrease between T$_{smp}^{on}$ and T$_p^{on}$. It
is only above T$_p^{on}$ that the J$_c$(T) values for ZFC and FCW
modes start to increase.

\section{Summary}

To summarize, we have investigated the effect of drive and disorder on
supercooling/superheating effects across the anomalous variations in
critical current density in several single crystals of 2H-NbSe$_2$ and
compared their results with those in a single crystal of CeRu$_2$. We
have mapped out in typically weakly pinned crystals of 2H-NbSe$_2$ and
CeRu$_2$ the phase coexistence regime of the stronger and the weaker
pinned pockets, where metastability effects manifest in a prominent
manner. The termination of the phase coexistence regime at higher
field and higher temperature end can be conveniently located in the
third harmonic ac susceptibility data. At lower fields ($\leq$ 1
kOe), the region I of the phase coexistence regime, in which the order
dominates, can be seen to continuously connect to the so called
reentrant disordered phase \cite{11,17}, where the intervortex spacing
presumably exceeds the range of interaction (a$_0$ $>$ $\lambda$,
where $\lambda$ is the penetration depth). Bitter decoration \cite{41}
and simulation studies \cite{42} have shown that the reentrant
disordered phase comprises polycrystalline form of flux line
lattice. So long as the domains are large, with radial correlation
length R$_c$ $\gg$ a$_0$, the weaker pinned regions dominate. The
electromagnetic response of the sample and the imposition of the
external driving forces can shrink the stronger pinned domain wall
regions. However, as the domain sizes shrink on crossing over to the
region II of the phase coexistence regime, the external driving forces
aid the process of complete amorphization of the vortex matter. The
vortices remain pinned in the amorphous region between T$^*$ and T$_c$
values in the typically weakly pinned crystals of low temperature
superconductors.

\section*{Acknowledgments}

We have benefited from discussions with E. Y. Andrei, S. Bhattacharya,
C. V. Tomy, A. Tulapurkar, D. Pal and D. Jaiswal-Nagar. We thank
P. L. Gammel for the crystal Y$^{\prime}$ of 2H-NbSe$_2$ and Y. Onuki for
the CeRu$_2$ crystal. We also thank R. S. Sannbhadti and U. V. Vaidya
for technical assistance. One of us (ADT) would like to acknowledge
the TIFR Endowment Fund for Kanwal Rekhi Career Development support.
\\

$^{\star}$ Email: ajay@tifr.res.in~~;~~ $^{\dagger}$ Email: satyajit@iitk.ac.in

\begin{figure} 
\includegraphics[scale=1.5,angle=0]{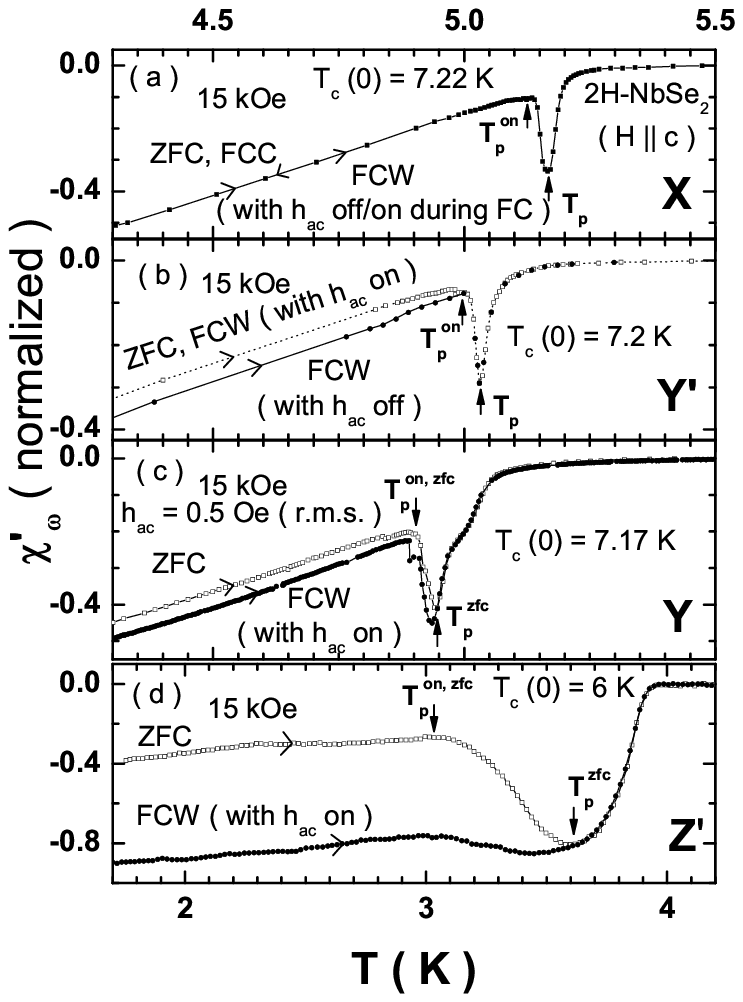}
\caption{In-phase ac susceptibility data showing peak effect (PE) in a field of 15 kOe with different thermomagnetic histories in single crystals of $2H$-$NbSe_2$. The crystals $X$, $Y^{\prime}$, $Y$ and $Z^{\prime}$ have progressively enhanced pinning. The crystal $Z^{\prime}$ has a lower $T_c(0)$ value of $\sim 6 K$ as it presumably contains 200 ppm of Fe impurity. The panel (a) shows that in the most weakly pinned crystal X, all the four responses, ZFC, FCC, FCW (recorded after $h_{ac}$ {\it on} or {\it off} during field cooling) are identical. In panel (b), the two FCW responses (with $h_{ac}$ {\it on} and {\it off} during field cooling) in the crystal $Y^{\prime}$ are different. While the former coincides with ZFC (dotted curve), the latter overlaps with FCC (the solid line). The panel (c) shows that in the crystal Y, the  $\chi_{\omega}^{\prime}$ response in the FCW with $h_{ac}$ {\it on} differs from that in the ZFC state. In the panel (d), the  $\chi_{\omega}^{\prime}$ data in the crystal $Z^{\prime}$ is such that the FCW response represents the supercooling of the vortex state existing at the peak position of the peak effect. The onset ($T_p^{on}$) and peak temperatures ($T_p^{zfc}$) of the PE in the ZFC mode have been identified in each of the panels.}
\end{figure}
\begin{figure} 
\includegraphics[scale=1.2,angle=0]{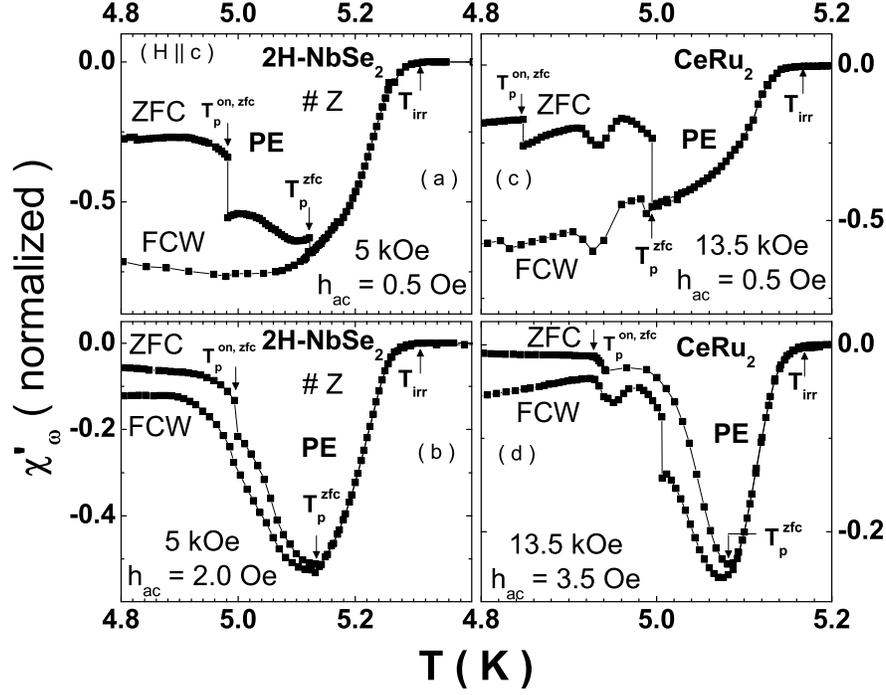}
\caption{Comparison of the  $\chi_{\omega}^{\prime}$ data at the fields indicated in the ZFC and the FCW modes with different amplitudes of the ac driving force ($h_{ac}$) in a typically weakly pinned single crystal Z of $2H$-$NbSe_2$ ($T_c(0) \approx 6.0 K$) and in a crystal of $CeRu_2$ ($T_c(0) \approx 6.3 K$). The positions of the onset and peak temperatures have been identified in each of the panels.}
\end{figure}
\begin{figure} 
\includegraphics[scale=1.5,angle=0]{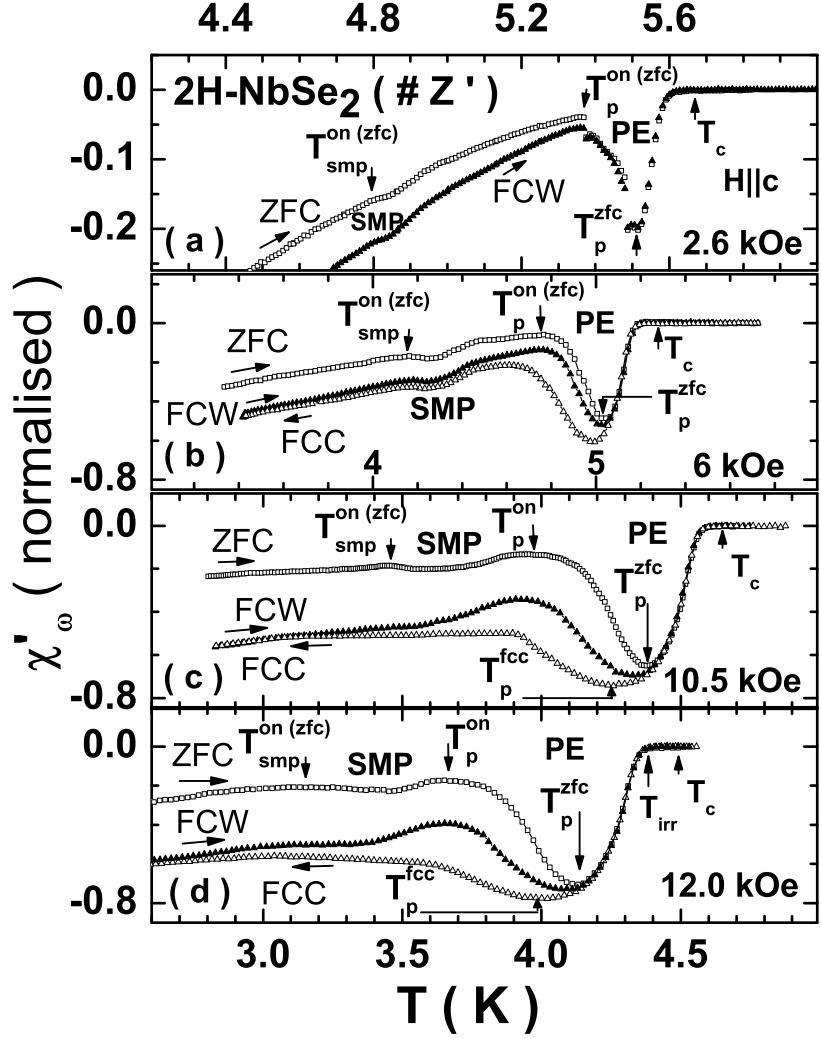}
\caption{ $\chi_{\omega}^{\prime}$ data at different applied fields for different thermomagnetic histories in a crystal $Z^{\prime}$ ($T_c(0) \approx 6.0 K$) of $2H$-$NbSe_2$. Note the occurrence of two anomalous variations corresponding to the SMP and the PE in each of the panels. The  $\chi_{\omega}^{\prime}(T)$ curves at H = 2.6 kOe in the panel (a) display the notion of stepwise fracturing across the PE, as in the Fig. 2 (a) for the sample Z at 5 kOe. The onset temperature of the SMP and the PE and the peak position of the PE in the ZFC mode have been marked in all the panels.}
\end{figure}
\begin{figure} 
\includegraphics[scale=1.5,angle=0]{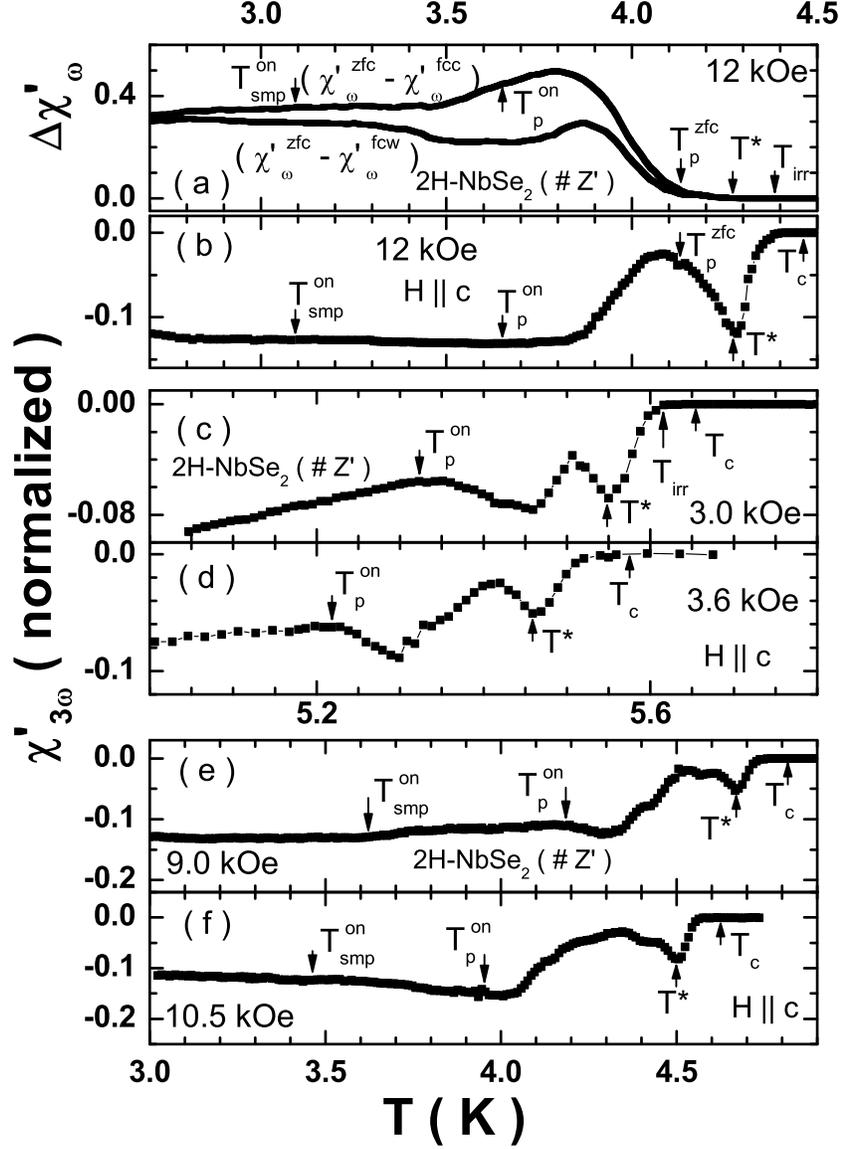}
\caption{The panels (a) and (b) show the plots of difference susceptibility $\Delta \chi_{\omega}^{\prime}$ i.e., ( $\chi_{\omega}^{\prime ~~ zfc}$ - $\chi_{\omega}^{\prime ~~ fcc}$) and ( $\chi_{\omega}^{\prime ~~ zfc}$ -  $\chi_{\omega}^{\prime ~~ fcw}$), and the third harmonic signal  $\chi_{3 \omega}^{\prime}$ in a field of 12 kOe ($H \| c$) in the crystal $Z^{\prime}$ of $2H$-$NbSe_2$. The positions of $T_{smp}^{on}$, $T_p^{on}$ and $T_p$ for the ZFC mode, as evident from the curve in Fig. 3 (d), have been identified in these two panels. $T^{\star}$ in the panel (a) identifies the temperature at which the difference ($\chi_{\omega}^{\prime ~~ zfc}$ - $\chi_{\omega}^{\prime ~~ fcc}$) vanishes and merges into the baseline. This value of $T^{\star}$ is then marked in the panel (b). Note that  $\chi_{3 \omega}^{\prime}$ response shows a monotonic decrease at $T > T^{\star}$. The panels (c) to (f) show  $\chi_{3 \omega}^{\prime}$ curves at different applied fields with $h_{ac} = 2.5 Oe$ (r.m.s.) in the crystal $Z^{\prime}$ of $2H$-$NbSe_2$. The respective $T^{\star}$ values have been identified in these panels. It is apparent that the  $\chi_{3 \omega}^{\prime}$ collapses above $T^{\star}$ and merges into the baseline as $T \rightarrow T_{irr} (< T_c)$, where $T_irr$ is the notional irreversibility temperature for a given hac and the frequency value in which the ac susceptibility measurements stand performed.}
\end{figure}
\begin{figure} 
\includegraphics[scale=1.2,angle=0]{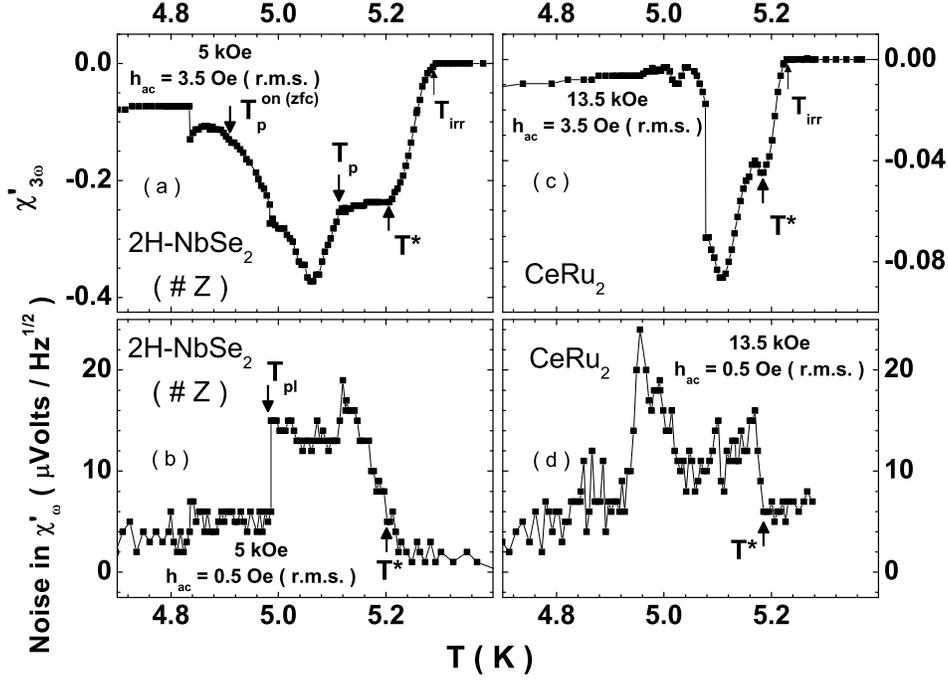}
\caption{A comparison of third harmonic, $\chi_{3 \omega}^{\prime}$ response and the noise signal in the ac susceptibility at the fields indicated in the typically weakly pinned single crystals of  $2H$-$NbSe_2$ ($\sharp Z$) and $CeRu_2$. The  $\chi_{3 \omega}^{\prime}$ data are recorded in $h_{ac}$ of 3.5 Oe (r.m.s.), whereas the measurements on the noise in  $\chi_{\omega}^{\prime}$ [see Ref. 21] were made with $h_{ac} =$ 0.5 Oe (r.m.s.). The $T_p^{on (zfc)}$ value in the panel (a) corresponds to the said value identified in the Fig. 2 (b) for the same $h_{ac}$ of 3.5 Oe (r.m.s.). The $T_{pl}$ value in the panel (b) identifies the onset temperature of the PE in $h_{ac}$ of 0.5 Oe (r.m.s.) in the sample Z, following the nomenclature as in Ref. 21. $T_{pl}(H)$ values denote the boundary of elastic to plastic flow for the driven vortex matter. A comparison of data in panels (b) and (d) show that $T^{\star}$ values identify limiting temperatures above which  $\chi_{3 \omega}^{\prime}$ responses monotonically decrease and the noise signals in  $\chi_{\omega}^{\prime}$ recede to the respective baselines.}
\end{figure}
\begin{figure} 
\includegraphics[scale=1.2,angle=0]{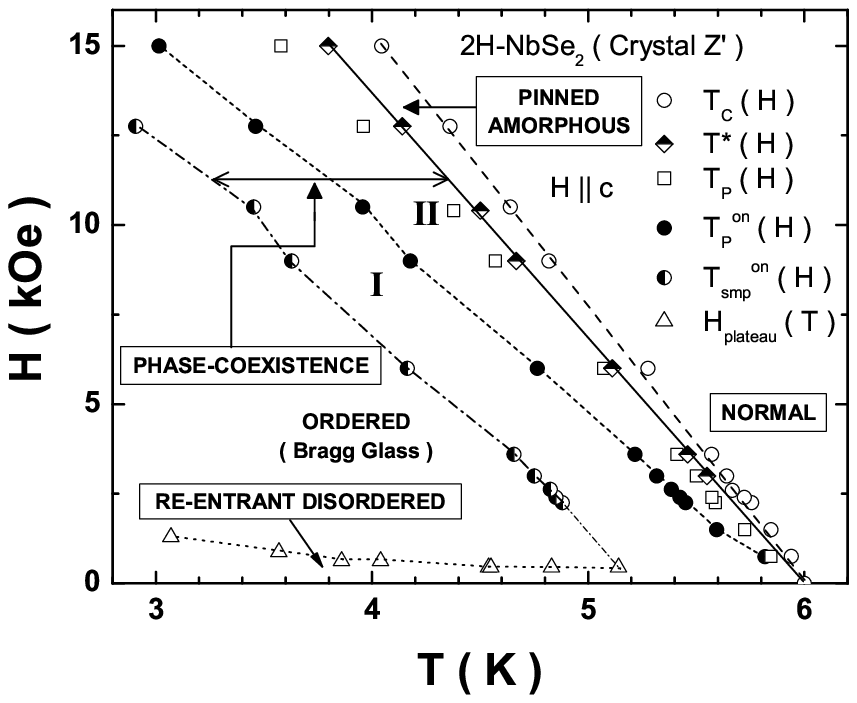}
\caption{Vortex phase diagram for $H \| c$ in a typically weakly pinned single crystal of $2H$-$NbSe_2$. The diagram comprises the values of onset temperatures of the second magnetization peak anomaly ($T_{smp}^{on}$) and the peak effect ($T_p^{on}$), the value of the peak temperature of the PE ($T_p^{zfc}$) in the ZFC mode, the limiting (spinodal) temperature $T^{\star}$, the $T_c(H)$ and the values of the crossover field, $H_{plateau}(T)$, determined from the normalized plots of critical current density versus normalized fields as in Ref. 21. The (H, T) phase space above $H_{plateau}(T)$ and below $T_{smp}^{on}(H)$ identifies the ordered Bragg (elastic) glass region. The intervening phase space between $T_{smp}^{on}(H)$ and $T^{\star}(H)$ comprises the co-existence regime of ordered (weaker pinned) and disordered regions. $T_p^{on}(H)$ values sub-divide the co-existence phase space into regions I and II in which ordered and disordered pockets dominate, respectively. The region I appear continuously connected to the phase space of the so called reentrant disordered regime, in which vortices are in the small bundle collective pinning regime. The phase space intervening between $T^{\star}$ and $T_c$ identifies the pinned amorphous region.}
\end{figure}


\begin{thebibliography}{}

\bibitem{1}
P. W. Anderson, in Basic Notions of Condensed Matter Physics, Addison-Wesley Publication Company, USA (1983).
\bibitem{2}
M. Tinkham, Introduction to Superconductivity, 2nd ed., McGraw-Hill International Edition (1996).
\bibitem{3}
G. Blatter, M. V. Feigel'man, V. B. Geshkenbein, A. I. Larkin, and V. M. Vinokur, {\it Rev. Mod. Phys.} {\bf 66}, 1125-1388 (1994) and references therein.
\bibitem{4}
M. Marchevsky, M. J. Higgins, and S. Bhattacharya, {\it Nature} {\bf 409}, 591 (2001).
\bibitem{5}
G. Blatter, V. B. Geshkenbein, and J. A. G. Koopmann, {\it Phys. Rev. Lett.} {\bf 92}, 067009 (2004) and references therein.
\bibitem{6}
A. B. Pippard, {\it Philos. Mag.} {\bf 19}, 217 (1969).
\bibitem{7}
A. I. Larkin, and Yu. N. Ovchinnikov, {\it J. Low Temp. Phys.}, {\bf 34}, 409 (1979).
\bibitem{8}
W. K. Kwok, J. A. Fendrich, C. J. van der Beek, and G. W. Crabtree, {\it Phys. Rev. Lett.} {\bf 73}, 2614 (1994).
\bibitem{9}
M. J. Higgins and S. Bhattacharya, {\it Physica} {\bf C257}, 232 (1996) and references therein.
\bibitem{10}
A. Soibel, E. Zeldov, M. Rappaport, Y. Myasoedov, T. Tamegai, S. Ooi, M. Konczykowski, V. B. Geshkenbein, {\it Nature} {\bf 406}, 282 (2000).
\bibitem{11}
S.S. Banerjee, T.V.C. Rao, A.K. Grover, M.J. Higgins, G.I. Menon, P.K. Mishra, D. Pal, S. Ramakrishnan, G. Ravikumar, V.C. Sahni, S. Sarkar, and C.V. Tomy, {\it Physica} {\bf C355}, 39 (2001).
\bibitem{12}
A. I. Larkin and Yu. V. Ovchinnikov, {\it Sov. Phys. JETP} {\bf 38}, 854 (1974).
\bibitem{13}
T. Giamarchi, and P. Le Doussal, {\it Phys. Rev.} {\bf B52}, 1242 (1995).
\bibitem{14}
S. S. Banerjee, N. G. Patil, S. Ramakrishnan, A. K. Grover, S. Bhattacharya, P. K. Mishra, G. Ravikumar, T. V. Chandrasekhar Rao, V. C. Sahni, M. J. Higgins, C. V. Tomy, G. Balakrishnan, and D. Mck. Paul, {\it Phys. Rev.} {\bf B59}, 6043 (1999).
\bibitem{15}
S. B. Roy and P. Chaddah, {\it J. Phys.: Condens. Matter} {\bf 9}, L625 (1997); ibid {\bf 10}, 4885 (1998).
\bibitem{16}
G. Ravikumar, P. K. Mishra, V. C. Sahni, S. S. Banerjee, A. K. Grover, S. Ramakrishnan, P. L. Gammel, D. J. Bishop, E. Bucher, M. J. Higgins, S. Bhattacharya, {\it Phys. Rev.} {\bf B61}, 12490 (2000).
\bibitem{17}
S. S. Banerjee, S. Ramakrishnan, A. K. Grover, G. Ravikumar, P. K. Mishra, V. C. Sahni, C. V. Tomy, G. Balakrishnan, D. Mck. Paul, P. L. Gammel, D. J. Bishop, E. Bucher, M. J. Higgins, and S. Bhattacharya, {\it Phys. Rev.} {\bf B62}, 11838 (2000).
\bibitem{18}
A. M. Troyanovski, M. van Hecke, N. Saha, J. Aarts, and P. H. Kes, {\it Phys. Rev. Lett.} {\bf 89}, 147006 (2002).
\bibitem{19}
X. S. Ling, S. R. Park, B. A. McClain, S. M. Choi, D. C. Dender, and J. W. Lynn, {\it Phys. Rev. Lett.} {\bf 86}, 712 (2001).
\bibitem{20}
Z. L. Xiao, O. Dogru, E. Y. Andrei, P. Shuk, and M. Greenblatt, {\it Phys. Rev. Lett.} {\bf 92}, 227004 (2004).
\bibitem{21}
S. S. Banerjee, N. G. Patil, S. Saha, S. Ramakrishnan, A. K. Grover, S. Bhattacharya, G. Ravikumar, P. K. Mishra, T. V. Chandrasekhar Rao, V. C. Sahni, M. J. Higgins, E. Yamamoto, Y. Haga, M. Hedo, Y. Inada, and Y. Onuki, {\it Phys. Rev.} {\bf B58}, 995 (1998).
\bibitem{22}
S. Ramakrishnan, S. Sundarum, R. S. Pandit and G. Chandra, {\it J. Physics E: Sci. Instrum.}, {\bf 18}, 650 (1985).
\bibitem{23}
S. S. Banerjee, Ph. D Thesis, 2000, University of Mumbai, Mumbai, India.
\bibitem{24}
S. Sarkar, D. Pal, S. S. Banerjee, S. Ramakrishnan, A. K. Grover, C. V. Tomy, G. Ravikumar, P. K. Mishra, V. C. Sahni, G. Balakrishnan, D. McK. Paul, and S. Bhattacharya, {\it Phys. Rev.} {\bf B61}, 12394 (2000).
\bibitem{25} 
X. S. Ling and J. I. Budnick, in Magnetic Susceptibility of Superconductors and other Spin Systems, edited by R. A. Hein, T. L. Francavilla, and D. H. Leibenberg (Plenum Press, New York, 1991), p. 377.
\bibitem{26}
Z. L. Xiao, E. Y. Andrei, P. Shuk, and M. Greenblatt, {\it Phys. Rev. Lett.} {\bf 86}, 2431 (2001).
\bibitem{27}
M. Steingart, A.G. Putz, and E.J. Kramer, {\it J. Appl. Phys.} {\bf 44}, 5580 (1973).
\bibitem{28}
W. Henderson, E. Y. Andrei, M. J. Higgins, and S. Bhattacharya, {\it Phys. Rev. Lett.} {\bf 77}, 2077 (1996).
\bibitem{29}
S. Sarkar, D. Pal, P. L. Paulose, S. Ramakrishnan, A. K. Grover, C. V. Tomy, D. Dasgupta, B. K. Sarma, G. Balakrishnan, and D. McK. Paul, {\it Phys. Rev.} {\bf B64}, 144510 (2001).
\bibitem{30}
C. P. Bean, {\it Phys. Rev. Lett.} {\bf 8}, 250 (1962).
\bibitem{31}
C. P. Bean, {\it Rev. Mod. Phys.} {\bf 36}, 31 (1964).
\bibitem{32}
L. Ji, H. Sohn, G. C. Spalding, C. J. Lobb and M. Tinkham, {\it Phys. Rev.} {\bf40}, 10936 (1989).
\bibitem{33}
Y. Yeshurun, M. W. McElfresh, A. P. Malozemoff, J. Hagerhorst-Trewhella, J. Mannhart, F. Holtzberg, and G. V. Chandrashekhar , {\it Phys. Rev.} {\bf B 42}, 6322 (1990).
\bibitem{34}
P. Chaddah, S. B. Roy, Shailendra Kumar and K. V. Bhagwat, {\it Phys. Rev.} {\bf B46}, 11737 (1992).
\bibitem{35}
Shailendra Kumar, S. B. Roy, P. Chaddah, Ram Prasad and N. C. Soni, {\it Physica} {\bf C191}, 450 (1992).
\bibitem{36}
G. Ravikumar, V. C. Sahni, A. K. Grover, S. Ramakrishnan, P. L. Gammel, D. J. Bishop, and E. Bucher, M. J. Higgins, S. Bhattacharya, {\it Phys. Rev.} {\bf B63}, 024505 (2001).
\bibitem{37}
A. C. Marley, M. J. Higgins and S. Bhattacharya, {\it Phys. Rev. Lett.} {\bf 74}, 3029 (1995).
\bibitem{38}
R. Merithew, M. W. Rabin, M. B. Weissman, M. J. Higgins and S. Bhattacharya, {\it Phys. Rev. Lett.} {\bf 77}, 3197 (1996).
\bibitem{39}
D. Li and B. Rosenstein, {\it Phys. Rev.} {\bf B65}, 220504 (2002); cond-mat/0305258; E. Y. Andrei, Z. L. Xiao, W. Henderson, Y.Paltiel, E. Zeldov, M. J. Higgins, S. Bhattacharya, P.Shuk and M. Greenblatt, Condensed Matter Theories, Nova Science Publishers, New York, USA, vol 16, p241-252 (2001).
\bibitem{40}
A. D. Thakur, S. S. Banerjee, M. J. Higgins, S. Ramakrishnan, and A. K. Grover, {\it Phys. Rev.} {\bf B72}, 134524 (2005).
\bibitem{41}  
M. Menghini, Y. Fasano, and F. de la Cruz, {\it Phys. Rev.} {\bf B65}, 064510 (2002).
\bibitem{42}
M. Chandran, R. T. Scalettar, and G. T. Zimányi, {\it Phys. Rev.} {\bf B69}, 024526 (2004).
 
\end{thebibliography}
\end{document}